\documentclass{sig-alternate-2013}

\usepackage{hyperref}
\usepackage{setspace}
\usepackage{graphicx}
\usepackage[space]{grffile}
\usepackage{latexsym}
\usepackage{amsfonts,amsmath,amssymb}
\usepackage{url}
\usepackage{subfigure}
\usepackage[utf8]{inputenc}
\usepackage{fancyref}
\hypersetup{colorlinks=false,pdfborder={0 0 0},}

\newfont{\mycrnotice}{ptmr8t at 7pt}
\newfont{\myconfname}{ptmri8t at 7pt}
\let\crnotice\mycrnotice%
\let\confname\myconfname%

\usepackage[ruled,lined,commentsnumbered]{algorithm2e}

\newcommand{\squishlist}{
  \begin{list}{$\bullet$}
   {
     \setlength{\itemsep}{0pt}
     \setlength{\parsep}{3pt}
     \setlength{\topsep}{3pt}
     \setlength{\partopsep}{0pt}
     \setlength{\leftmargin}{1.5em}
     \setlength{\labelwidth}{1em}
     \setlength{\labelsep}{0.5em} } }

\newcommand{\squishlisttwo}{
  \begin{list}{$\bullet$}
   {
     \setlength{\itemsep}{0pt}
     \setlength{\parsep}{0pt}
     \setlength{\topsep}{0pt}
     \setlength{\partopsep}{0pt}
     \setlength{\leftmargin}{2em}
     \setlength{\labelwidth}{1.5em}
     \setlength{\labelsep}{0.5em} } }

\newcommand{\squishdef}{
  \begin{list}{}
   {
     \setlength{\itemsep}{0pt}
     \setlength{\parsep}{3pt}
     \setlength{\topsep}{3pt}
     \setlength{\partopsep}{0pt}
     \setlength{\leftmargin}{1.5em}
     \setlength{\labelwidth}{1em}
     \setlength{\labelsep}{0.5em} } }

\newcommand{\squishend}{
   \end{list}  }

\begin{document}
%
\permission{Permission to make digital or hard copies of all or part of this work for personal or classroom use is granted without fee provided that copies are not made or distributed for profit or commercial advantage and that copies bear this notice and the full citation on the first page. Copyrights for components of this work owned by others than ACM must be honored. Abstracting with credit is permitted. To copy otherwise, or republish, to post on servers or to redistribute to lists, requires prior specific permission and/or a fee. Request permissions from permissions@acm.org.}
\conferenceinfo{WSDM'15,}{February 2--6, 2015, Shanghai, China.}
\copyrightetc{Copyright 2015 ACM \the\acmcopyr}
\crdata{978-1-4503-3317-7/15/02\ ...\$15.00.\\
http://dx.doi.org/10.1145/2684822.2685296}

\clubpenalty=10000 
\widowpenalty = 10000

\title{On the Accuracy of Hyper-local Geotagging \\of Social
Media Content}

%
%
%
%
%

\numberofauthors{1} 
\author{
\alignauthor David Flatow$^{1,2}$, Mor Naaman$^1$, Ke Eddie Xie$^{1,3}$, Yana Volkovich$^{1,4}$, Yaron Kanza$^{1,5}$ \\
     \affaddr{ $^1$ Jacobs Institute, Cornell Tech,  111 8th Ave New York, NY 10011, USA } \\
     \affaddr{ $^2$ Stanford University, 450 Serra Mall, Stanford, CA 94305, USA} \\
     \affaddr{ $^3$ Twitter Inc., 1355 Market Street, San Francisco, CA 94103, USA } \\
     \affaddr{ $^4$ Barcelona Media, Av. Diagonal 177, 08018 Barcelona, Spain} \\
     \affaddr{ $^5$ Technion -- Israel Institute of Technology, Haifa 36000, Israel}\\
    	 \email{ dflatow@stanford.edu,  \{mor, kanza, yana\}@jacobs.cornell.edu, exie@twitter.com}
     }

\maketitle

\begin{abstract}
Social media users share billions of items per year, only a small
fraction of which is geotagged. We present a data-driven approach for
identifying non-geotagged content items that can be associated with a
\emph{hyper-local} geographic area by modeling the location
distributions of $n$-grams that appear in the text. We explore
the trade-off between accuracy and coverage of this method.
Further, we explore differences across content received from multiple
platforms and devices, and show, for example, that content shared via different sources and applications produces significantly different geographic distributions, and that it is preferred to model and predict location for items according to their source. Our findings show the potential and the
bounds of a data-driven approach to assigning location data to short social media texts, and
offer implications for all applications that use data-driven approaches
to locate content.

\end{abstract}

%
\bibliographystyle{abbrv}
%
%

\category{H.3.5}{Information Storage and Retrieval}{On-line Information Services}[Web-based services]

\keywords{Geotagging; Social Media; Location-based Services}

\section{Introduction}\label{sec_intro}
The vast amounts of data shared on social media reflect people's attitudes, attention, activities and interests, thus offering unique opportunities to analyze and reason about our world and our society. With associated geographic information, these social-media items allow us to understand, for the first time, what geographic areas people are paying attention \textit{to}, and where they pay attention \textit{from}. Mining this dataset can prove hugely valuable to a diverse set of applications, including improving city management \cite{Xia:2014:CRS:2567948.2577020}, journalism \cite{matias2014newspad}, tourism \cite{Crandall_2009,Rattenbury_2007,Quercia2014}, health~\cite{sadilek2012modeling} and more.  

We call a social media item \textit{geotagged} when it is associated with geographic coordinates, usually indicating where the item was created. However, only a minor portion of the content posted on social media sites such as Twitter, Instagram and Flickr is geotagged. Reported and estimated numbers range from 2\% of the items for Twitter\footnote{\tiny http://blog.gnip.com/twitter-geo-data-enrichment/}, to 25\% on Instagram\footnote{\tiny http://bits.blogs.nytimes.com/2012/08/16/instagram-refreshes-app-include-photo-maps/}.

Nevertheless, many of the items that are \textit{not} geotagged may still provide valuable geographic information if they can be associated correctly with the location where they were created.
%
%
In this work, we are interested in associating non-geotagged social media items with \emph{hyper-local} geographic locations. Such a process will increase the amount of data associated with a location (e.g., a park, a venue) and allow for more robust search and data mining applications. For example, increasing the amount of content available for Madison Square Park in New York may allow park administrators to more robustly model and monitor activities using public social media data. 

Most recent work on locating non-geotagged content in social media focuses on inferring locations of users~\cite{Chandra,Cheng_2010,Kinsella_2011} rather than of individual content items. A common approach to the problem is identifying spatial aspects of phrases in unstructured texts (e.g.,~text in items posted by the user, or the text in the user's profile). While attempting to improve geographic coverage, these systems, for the most part, do not consider accuracy bounds, and instead emphasize the extraction of a general ``user location''. For example, posting about the Steelers (a Pittsburgh football team) could increase a user's probability of being located in Pittsburgh, but would not be able to expose where in Pittsburgh the user may be. Conversely, our goal here is to identify individual social media items that can be located with high precision inside a \textit{small (hyper-local) geographic region}. In particular, this paper investigates a data-driven approach for hyper-localization of content, and explores the bounds and trade-offs of such a method.

Our approach to localizing social media items involves: (1)~identifying phrases that can be accurately (based on data) associated with a specific location, and (2)~identifying items that contain these phrases. 
In an approach inspired by Priedhorsky et al.~\cite{Priedhorsky_2014}, we train a model on text contained in geotagged items with the goal of identifying $n$-grams that are geo-specific: a \textit{large} portion of the items containing the $n$-gram are posted from a \textit{small} area (the portion of items and the area size are configurable parameters).
Our model generates an iterative Gaussian model for each $n$-gram in order to discover hyper-local phrases that can be used to predict locations for non-geotagged content.

 
Indeed, localizing social media items is not an easy task for several reasons~\cite{Zuyev2014}. First, it is a priori unknown which areas will be associated with textual terms and which will not. Second, it is a priori unknown which $n$-grams will be associated with hyper-local regions. Thus, a naive search by examining areas or $n$-grams does not work well. More importantly, there are many terms that can be roughly associated with areas, but are not localized enough to be associated  with a hyper-local region, and erroneously using such terms may lead to errors or inaccurate results. Thus, it is important to provide an accurate localization method but also understand the limitation of the proposed method.

We explore the bounds, properties and trade-offs of such a hyper-local geotagging solution, on (and across) different data sources. An important question this paper addresses is whether these geographic models are specific to the type or source of content. For example, are location models discovered based on posts of iPhone users also relevant for localizing posts of Android users? Can Twitter tweets be used for localizing Instagram photos or vice versa? Answering these questions could be critical for data mining applications that perform and build on geotagged data.

We collected more than 14 million Twitter posts that were geotagged by users within the area of New York City. The Twitter posts in the dataset were created using different devices (iPhone, Android) and originated from different applications (Instagram, Foursquare). We use these data for our training and test sets. New York is one of the most georeferenced cities in the world (it covers around 2.5\% of the total geotagged Twitter content~\cite{Leetaru_2013}), and serves as an excellent testing ground for our methods. Of course, the framework would be effective in other geographical regions as well. 

The contributions of this work include:
\squishlist
\item Introducing a data-driven approach to identify phrases ($n$-grams) associated with hyper-local regions;
\item Investigating and evaluating the approach across multiple social media data types and data sources;   
\item Exposing the limitations and properties of models from different sources and at different scales;  
\squishend

\section{Related work}\label{sec_relwork}
The problem of geotagging social Web data has received significant attention recently. Related studies can be divided into three broad (and overlapping) areas: geotagging of social media content, understanding characteristics of geographical areas from social media data, and modeling textual location references in social media. 

\subsection{Geotagging Social Content}\label{sec_relwork_geotagging}
Most recent studies on automatic text-based geolocation of social media content aimed to identify location(s) of a social media \textit{user}. While localizing users or posts are related problems, they have distinct properties and biases. Knowing users' overall location is, perhaps, a first step to predicting location for individual posts. For Twitter, research has shown that \emph{home}, or primary, location for each user can be learned by analyzing content of tweets. These predictions usually do not go below city-level~\cite{Chandra,Cheng_2010,Compton14,Kinsella_2011,Mahmud_14}. A few recent works incorporated additional features such as time zones~\cite{Mahmud_14} or friends locations~\cite{Compton14} to find correctly the location of the home city, for roughly 80\% of Twitter users, with median error of 6.33km. Schulz et al.~\cite{schulz2013multi} proposed multi-indicator approach that combines various spatial indicators from the user’s profile and the tweet’s message for estimating the location of a tweet as well as the user’s home location. Here, in contrast, we are not interested in a user's overall location. Our goal is identifying content items that were posted from specific geographic areas, with finer granularity than city-level. 

Some studies applied traditional language models to geotagging, e.g. Hong et al.~\cite{Hong_2012} used $k$-means clustering, Eisenstein et al.~\cite{eisenstein2011sparse} used Dirichlet Process mixture, and Kling et al.~\cite{kling2014detecting} used multi-Dirichlet process. Our work, however, is related to the methods introduced in~\cite{Priedhorsky_2014} and~\cite{Hau_wen_Chang}. Priedhorsky et al.~\cite{Priedhorsky_2014}, for example, used a Gaussian Mixture Model (GMM) to estimate a tweet's location based on the distribution of $n$-grams that appear in the tweet and associated content to it (e.g. user profile information). In particular, they generated geographic density estimates for all $n$-grams, and used density information to provide a final location estimate for content, regardless of the geographic scope. In contrast, we attempt to identify $n$-grams that can predict the location of a tweet with high precision. 

Mapping social media content to geographical locations typically implies some discretization of the spatial area. For example, geographical locations might be clustered as a grid \cite{Serdyukov_2009,Wing_2011}. However, the fixed-grid based representations have a limitation of not capturing variability in shapes and sizes of geographical regions.  One of the possible ways to overpass such limitation is to define an alternative grid construction, for example by using $k$-d trees~\cite{roller2012supervised}. A different way of representing geographic areas is to use a continuous function over the space~\cite{Priedhorsky_2014}, an approach we take in this work as well.

\subsection{Characterizing Geographic Areas}\label{sec_relwork_locations}
A related set of studies used information about geographic regions in geotagged social media to extract information and characterize geographic areas ~\cite{Ahern_2007,Crandall_2009,dalvi2012object,lian2014mining,Quercia_2012,Thomee_2013}.

Ahern et al.~\cite{Ahern_2007} proposed a model that aggregates knowledge in the form of ``representative tags'' for arbitrary areas in the world by analyzing tags associated with the geo-referenced Flickr images. Crandell et al.~\cite{Crandall_2009} used Flickr to find relations between photos and popular places in which the photos were taken and showed how to find representative images for popular landmarks. Similarly, Kennedy et al.~\cite{kennedy2008generating} generate representative sets of images for landmarks using Flickr data. Quercia et al.~\cite{Quercia_2012} proposed applying sentiment analysis to geo-referenced tweets in London in order to find the areas of the city characterized by ``well being''. A recent review by Tasse~\cite{tasseusing} listed other possible applications of social media for understanding urban areas.


\subsection{Characterizing Location References}\label{sec_relwork_loc_ref}
Efforts were made to characterize location references (in text) within social media content. 
Rattenbury et al.~\cite{Rattenbury_2007} used an approach similar to the one presented in this paper, when trying to identify Flickr tags that refer to specific geographic places or specific events based on the spatiotemporal distribution of geotagged photos carrying that tag. However, the authors did not apply their models to geocoding new items, did not explore hyper-local content, and did not extract phrases from free-form text like we do here. Further, Thomee and Morales~\cite{Thomee_2014} find that different language variants of toponyms can be mapped to each other by exploiting the geographic distribution of tagged media.

Recent work by Shaw et al.~\cite{Shaw_2013} mapped users noisy check-ins on Foursquare to semantically extract meaningful suggestions from a database of known points of interest. In particular, by aggregating locations from geotagged check-ins, the authors were able to create geographic models for different venues using multi-dimensional Gaussian models. Earlier work from Flickr, Alpha Shapes\footnote{code.flickr.net/2008/10/30/the-shape-of-alpha}, modeled information available from geotagged images on Flickr to create geographic models for places like neighborhoods, towns, etc. Finally, Li et al.~\cite{Li_2014} not only explored point-of-interest mentions on Twitter but also connected them to the relative temporal values of the visits.

\section{Geotagging Framework} \label{sec_framework}
Our framework for associating geographic locations with social media items uses \textit{training data} to identify $n$-grams that are \textit{geo-specific} --- $n$-grams whose associated items' locations have little geographic variance. The process also results in location estimates for the $n$-grams that are deemed geo-specific. The discovered $n$-grams are used for geotagging items from the \textit{test set}, where items are associated with a location based on the $n$-grams they contain. For example, if ``Madison Square Park'' is detected as a geo-specific $n$-gram, a tweet in the test set that includes such $n$-gram will be associated with the location assigned to this $n$-gram. 

Note that the method we use does not aim to produce a general ``best guess'' for the location of an item; nor does it aim to identify an approximate location of a user, given a set of items. Instead, we are interested in identifying individual items that, with certain (high) accuracy, can be associated with a hyper-local location, such as a neighborhood, landmark or street corner. Note that multi-item and user-level information may very well be useful even in this scenario. We discuss this opportunity in more depth below.

Next, we describe the statistical process of identifying a geo-specific $n$-gram and the procedure for modeling its location. We detail the procedure for assigning locations to tweets in the test set (Section~\ref{sec_predicting_location}). Finally, we discuss and propose metrics (Section~\ref{sec_metrics}) to evaluate the performance, accuracy and bounds of this approach.

\subsection{Localizing $n$-grams} \label{sec_modeling_location}
We start by finding frequent $n$-grams, i.e.~$n$-grams that appear in many posts within the training dataset. Next, we associate each $n$-gram with the geographic locations of the posts containing it.
The main task is to use this data to (1)~decide whether a given $n$-gram is geo-specific, and (2)~model the location for a given geo-specific $n$-gram. We use a data-driven approach, inspired by Priedhorsky et al.~\cite{Priedhorsky_2014} and Chang et al.~\cite{Hau_wen_Chang}. In brief, we apply an iterative procedure of discovering the location model for a given $n$-gram, by removing outliers and recomputing a Gaussian model for the remaining locations in each step. For each $n$-gram $w_j$, this process determines whether locations of tweets associated with $w_j$ are describing a hyper-local area, and if so, computes the parameters of the area. 

For convenience, we refer to social media content items as tweets, i.e. posts on Twitter, in the definitions below. However, the methods we propose could be applied to other types of social media items, e.g. Instagram photos. We denote as $T$ the set of tweets in the training set and each tweet from this set as $t_i \in T$. The geographic location of $t_i$ is denoted $l_{i}$. An $n$-gram is a consecutive sequence of $n$ terms. A tweet contains an $n$-gram if the $n$ terms appear contiguously in the tweet. 

Let $T_{w_j}$ be the set of all tweets containing the $n$-gram $w_j$. Let $\bar{T}_{w_j}$ be some subset of $T_{w_j}$ and let $\bar{L}_{w_j}$ be the set of locations for tweets $t_i \in \bar{T}_{w_j}$. Further, we can fit a two dimensional Gaussian~$\mathcal{N}_j$ to the set of locations~$\bar{L}_{w_j}$ and define, based on the Gaussian, an ellipse $E_{2,\mathcal{N}_j}$. We construct $E_{2,\mathcal{N}_j}$ to have orientation, shape, and center defined by $\mathcal{N}_j$. Specifically, $E_{2,\mathcal{N}_j}$ is scaled so that its major and minor axes have lengths $2\sigma_1$ and $2\sigma_2$, where $\sigma_1^2$ and $\sigma_2^2$ are the first and second eigenvalues of $\Sigma$ respectively. We decide that $w_j$ is geo-specific if we find a subset $\bar{T}_{w_j} \subseteq T_{w_j}$, for adjustable parameters $s$ and $\tau$, such that: 
\squishlist
\item The fraction of tweets $\frac{|\bar{T}_{w_j}|}{|T_{w_j}|}$ is greater than a ratio parameter $\tau$, and
\item The area of the ellipse $E_{2,\mathcal{N}_j}$ is smaller than an area parameter $s$
\squishend

We define a characteristic function $\mathcal{X}(s, \tau, w_j)$ such that $\mathcal{X}(s, \tau, w_j) = 1$ if the algorithm decides that $w_j$ is geo-specific under parameters $(s, \tau)$. If $\mathcal{X}(s, \tau, w_j) = 1$, we output $\mathcal{N}_j$. Using this final Gaussian, $\mathcal{N}_j$, we can compute the ellipse $E_{2,\mathcal{N}_j}$ that approximates the area represented by $w_j$.

To find whether a subset $\bar{T}_{w_j}$ that matches this criteria exist, we apply and iterative modeling procedure, described in more detail in Algorithm~\ref{algo_iterations}. The procedure includes the following steps, starting with $\bar{T}_{w_j} = T_{w_j}$ (the full set of tweets):
\begin{enumerate}
\item If $\frac{|\bar{T}_{w_j}|}{|T_{w_j}|} < \tau$ , set $\mathcal{X}(s, \tau, w_j) = 0$ and break. 
\item Fit the two dimensional Gaussian $\mathcal{N}_j$ to the set of locations~$\bar{L}_{w_j}$ and compute $E_{2,\mathcal{N}_j}$.
\item If area($E_{2,\mathcal{N}_j}$) $\le s$, set $\mathcal{X}(s, \tau, w_j) = 1$ and break. 
\item Remove all the tweets outside of $E_{2,\mathcal{N}_j}$ (i.e.~are more than two standard deviations from the center of $\mathcal{N}_j$).
\end{enumerate}

We repeat this process until some $\bar{L}_{w_j}$  is deemed \textit{geo-specific}, the proportion of tweets goes below $\tau$, or an iteration limit is met. We say that $n$-gram $w_j$ is \textit{geo-specific} if at any point in an iterative outlier removal procedure both parameters $s$ and $\tau$ are satisfied. For example, if $s=4km^2$ and $\tau=0.8$, $w_j$ is geo-specific if at least 80\% of the posts in $T_{w_j}$ are contained in some $E_{2,\mathcal{N}_j}$ with an area smaller than $4km^2$ at any step in the iterative outlier removal process.


\begin{algorithm}[th]
\KwData{$L(w_j)$, \textit{MaxArea:} $s$, \textit{Ratio~threshold:} $\tau$, \textit{Iteration Limit:} $k$}

\KwResult{\textrm{Boolean: $GeoSpecific_j$; Gaussian: $\mathcal{N}_j$}}

$GeoSpecific_j\gets \textit{false}$\;
$\textit{iteration}\gets 0$\;
$\bar{L}_{w_j} \gets  L_{w_j}$\;
\While{$\textit{iteration}\leq k$}{
      \If{$\frac{|\bar{L}_{w_j}|}{|L_{w_j}|} < \tau$}{
        break\;
    }
  $\mu\gets \textit{mean}(\bar{L}_{w_j})$\;   
  $\Sigma\gets \textit{cov}(\bar{L}_{w_j})$\;
  $\mathcal{N}_j \gets \{\mu,\Sigma\}$\;
    \If{$\textit{Area}(E_{2,\mathcal{N}_j})\leq s$}{
	    $GeoSpecific_j\gets \textit{true}$\;
        break\;
    }
  $L_{w_j}^{\textit{temp}} \gets \emptyset$\;
    \For{$i$ \textbf{in} $\bar{L}_{w_j}$}{
	    \If{$i$ is in $E_{2,\mathcal{N}_j}$} {
		    $L_{w_j}^{\textit{temp}}.\textit{add}(i)$\;
	    }
    }
    $\bar{L}_{w_j} \gets L_{w_j}^{\textit{temp}}$\; 

    $\textit{iteration} \gets \textit{iteration}+1$\;
}
\textbf{return} $GeoSpecific_j,\mathcal{N}_j$\;
\caption{Iterative Modeling Procedure} \label{algo_iterations}
\end{algorithm}

\subsection{Assigning a Location to a Tweet}
\label{sec_predicting_location}

When associating locations for a tweet $t_i$ in the test set, we follow these simple rules. We first identify all $n$-grams $w_j$ such that $w_j$ is contained in $t_i$, $\mathcal{X}(s, \tau, w_j) = 1$ and $w_j$ is not contained in any other $n$-gram $w_k$ that satisfies the requirement. In other words, we find the longest possible $n$-grams in $t_i$ that are geo-specific. 

In our analysis, we associate a tweet with a location $\hat{l}_i$ according to the center of the Gaussian model for an $n$-gram it contains. If a tweet $t_i \in T_{\textit{test}}$ contains a single geo-specific $n$-gram $w_j$, we associate the tweet with $\mu_j$ from $\mathcal{N}(\mu_j,\Sigma_j)$. If a tweet contains more than one geo-specific $n$-gram whose centers are all pairwise within $0.5km$ of each other, we use the parameters of the most common of these n-grams. Setting the minimum ratio $\tau$ at 0.8 and the maximum allowed area $s$ fixed at $4km^2$, less than one percent of tweets with at least one geo-specific $n$-gram in the iPhone or Android test set contained multiple geo-specific $n$-grams. For the Instagram test sets, while roughly 27\% of the tweets in the test set contained multiple geo-specific $n$-grams, 73\% of these instances could be explained by multiple geo-specific $n$-grams that were nearby each other and of these, 42\% were simply cases where the $n$-grams were subsets of a longer phrase. For example, ``New York Public'' and ``York Public Library'' are subsets of the longer phrase ``New York Public Library''. Other approaches to handling multiple geo-specific $n$-grams include choosing the most accurate $n$-gram (the one with the smallest error) as the $n$-gram predicting the location of the tweet, using language processing to extract more information, or explicitly modeling the co-occurrence of such $n$-grams   (we discuss such approaches in Section~\ref{sec_concl}).

Finally, and naturally, a tweet $t_i$ that does not contain any geo-specific $n$-gram $w_j$ is not associated with a location.

%
%
\newpage
\subsection{Metrics}\label{sec_metrics}
We use three key metrics to evaluate the performance of our hyper-local geotagging framework on different datasets and with different parameters: error, precision, and coverage. In all cases, we have a training set of tweets with known locations, and a test set of tweets, with locations hidden, for which the algorithm decides whether or not to assign an estimated location. In this setup, the \textit{error\/} captures the geographic distance between an assigned location and a true location of a tweet from the set of test tweets. \textit{Precision\/} is the proportion of tweets whose true location is within the core ellipse of the tweet's assigned Gaussian. 
\textit{Coverage\/} refers to the portion of tweets in the test dataset for which the algorithm is able to assign a location. 

Next, we formalize these metrics. Note that we do not directly evaluate the $n$-gram information that is produced by the algorithm. One option would be to manually code the $n$-gram that the algorithm determines as geo-specific, e.g. by a human judge. Another option would be to evaluate the modeled locations associated with each $n$-gram, again, by a human. However, since we are taking a data-driven approach here, we do not need to directly evaluate the $n$-grams. In fact, the data may expose trends that would be non-obvious to a person. For example, the $n$-gram Nintendo turns out to be location-sensitive in some New York datasets \cite{Zuyev2014}, due to the Nintendo store at Rockefeller Plaza. 

\subsubsection{Error}\label{sec_error}
We define the error for tweet $t_i \in  T_{test}$ as the geographic distance between the true (hidden) location of the tweet $l_i$ and the tweet's estimated location $\hat{l}_i$. We use the center of the Gaussian associated with the tweet, as described in Section~\ref{sec_predicting_location}, as the estimated location. We use the Haversine distance $d(l_i, \hat{l}_i)$ to compute the error. 
Another option to define the error would be computing the distance from the ellipse $E_{2,\mathcal{N}_j}$ (not from the center) defined by the Gaussian model assigned to the tweet. We do not use this type of error measurement in this work, for simplicity. The {\em accuracy\/} is the inverse of the error. We consider it with respect to a given accuracy parameter $\delta$. The accuracy is equal to 1 when $d(l_i, \hat{l}_i)\leq \delta$, and it is $d(l_i, \hat{l}_i)^{-1}$ otherwise. Thus, when the error is small the accuracy is high (near 1) and when the error much larger than $\delta$, the accuracy is low.


\subsubsection{Precision}\label{sec_precision}
We define precision as the fraction of tweets whose true locations fall within the core ellipse $E_{2,\mathcal{N}_j}$ computed for the $n$-gram $w_j$. In other words, the precision for a test set $T_{test}$ of size $n$ is $\frac{\sum_{i=1}^n{R(t_i)}}{n}$, where $R(t_i)=1$ if $l_i$ is in $E_{2,\mathcal{N}_j}$, and $0$ otherwise. 
This method has the property that the criterion for a precise prediction is a function of parameter choices $s$ and $\tau$. For example, a prediction that is deemed precise in a model with large area $s$ (a model with a loose definition of hyper-locality) may not be precise in a model with a small area $s$ (stricter definition of hyper-locality).

\subsubsection{Covearge}\label{sec_coverage}
Coverage is defined as the fraction of tweets in the test set for which we can predict a location given a set of model parameters. More specifically, it is the fraction of tweets in the test set with a single geo-specific $n$-gram or multiple adjacent $n$-grams. Maximizing coverage is a conflicting goal to maximizing accuracy and precision. Setting the parameters such that only $n$-grams with highly accurate models are used (e.g. by setting a small area $s$ and high minimum ratio $\tau$) can result in a small number of $n$-grams that are geo-specific, and by extension, a small number of tweets that contain these $n$-grams.

\section{Experiments}\label{sec_experiment}


We explore the trade-offs and properties of the $n$-gram based geocoding approach in a series of experiments with a number of Twitter datasets. As mentioned above, even within Twitter, there are multiple sources and types of data shared by different users. For example, tweets can include photos or contain text-only; the source of the tweets can be another application, like the photo-sharing application Instagram or the location-check-in application Foursquare; or tweets can be generated by different types of devices, such as the Andriod phone or the iPhone. Research so far had not considered the source or type of content when creating location and geographic models. Here we explore that issue in more depth as we expect the source will have a significant effect on the geographic distribution of content. 


 
We describe the datasets and provide more detail on extracting candidate $n$-grams in Section~\ref{sec_data}. In Section~\ref{sec_parameters} we examine the performance of the method in respect to the metrics defined above, using the different datasets, and with different parameter settings. In Section~\ref{sec_cross_model} we examine the performance across datasets, or rather, what happens when we mix content from different sources in our training and testing datasets. 

\subsection{Datasets}\label{sec_data}
We collected geotagged tweets shared from the New York City area spanning two years from July, 2012 to July, 2014. This core dataset of 14.5 million geotagged tweets is the basis for the derived datasets we use in all experiments. We extract from the dataset four mutually exclusive sets of tweets according to each tweet's original source application: Foursquare, Instagram, Twitter for iPhone, and Twitter for Android (the source information is available for each tweet retrieved using the Twitter API). For all experiments, we use the first 651 days of tweets (July 21st, 2012 to May 3rd, 2014) as training data, and the remaining 80 days (May 4th to July 23rd, 2014) as test data with a one~day gap in-between to simulate a real-world case where you build models on extant data before applying to incoming information. The rate at which content is generated is different for each source. For example, roughly four times as many tweets with location information in the New York City area are shared from iPhones as there are from Android phones. In order to evaluate performance on a per training item basis we sample random tweets from each data source so that all training sets and all testing sets, respectively, contain the same number of items (training: 1,014,574, testing: 257,083). Finally, we created a TW-All dataset, where we randomly sampled from the entire set of Tweets to create a dataset that mimics the properties of the full dataset but is comparable in size to the others. The different datasets and distinct number of users in each of the sampled datasets we used for our experiments are listed in Table~\ref{table_datasets}.


The source applications are different on many --- sometimes interleaving --- dimensions, including their function, the device they are running on, and even the demographics of their users. Instagram is a photo-sharing application whose users can choose to post their photos to Twitter, often with a caption much shorter than a ``normal'' tweet. Foursquare is a location check-in app where users can choose to share their check-in (``I'm at Cornell Tech'') on Twitter, often with just the check-in default text. Finally, Twitter for iPhone and Twitter for Android are two Twitter mobile applications, that, while similar in nature and design, differ in the type of mobile device they run on, which subsequently also results in a different user population (note that we do not have device information for posts from Foursquare or Instagram). Given these differences, we expect the different sources to produce different types of information, and, therefore, different models of location. 



\begin{table}[t]
\scriptsize{
\begin{tabular}{|l|l|l|l|}
\hline \textbf{Name} & \textbf{\% of Total} & \textbf{\% Used} & \textbf{Users} \\
 & \textbf{In dataset} & \textbf{For train/test} & \textbf{In train/test} \\
\hline
\hline
TW-iPhone & 60\% & 15\% & 151,431 \\
\hline
TW-Android & 16\% & 57\% & 72,692 \\
\hline
TW-Instagram &  9\% & 97\% &166,965 \\
\hline
TW-Foursquare & 9\% & 100\% & 78,598 \\
\hline
TW-ALL & 100\% & 9\% & 222,608 \\
\hline
Other (not used) & 6\% & 0\% & N.A. \\
\hline
\end{tabular}
}
\caption{The datasets used in the experiments}
    \label{table_datasets}
\end{table}

The items in our test data, extracted from the same source as the modeling/training data, conveniently have associated location information. As a result, we can robustly evaluate our methods through experimentation. However, we do note that using this test data may also introduce bias. Of particular concern is that the distribution of text and locations in the global Twitter dataset are different than those in the geotagged tweets. However, we believe our approach is useful enough to create a baseline understanding of the potential (and challenges) of these types of methods. 

To generate candidate $n$-grams, we tokenized the message text and location description into $n$-grams of length $n=1,2,3$ by splitting words delimited by whitespace and removing  English stop words. In order to mitigate spam (often one user sharing hundreds of similar spam messages) $n$-grams used by fewer than 5 unique users or appearing in fewer than 20 unique tweets were removed. 

\subsection{Exploring the Parameter Space}\label{sec_parameters}
For each parameter setting in each dataset we examine the impact on various measures of performance. In particular, we vary the minimum ratio parameter $\tau$ and the area $s$, defined in Section~\ref{sec_modeling_location}, and examine the effect on accuracy, precision and coverage as defined in Section~\ref{sec_metrics}. Recall that the minimum ratio parameter $\tau$ controls the precision of the model, i.e. how much content we allow in the model that is outside the core ellipse,. The parameter $s$ controls the maximum size of the core ellipse for each model. Higher minimum ratios and smaller maximum areas correspond to stricter standards for classifying a given $n$-gram as geo-specific and generally lead to higher accuracy and lower coverage. To capture the balance between accuracy/precision and coverage we also include the F-score, $2\frac{Precision*Coverage}{Precision + Coverage}$, in the figures. 

\begin{figure}[h!]
\begin{center}
\subfigure{\includegraphics[width=0.8\columnwidth]{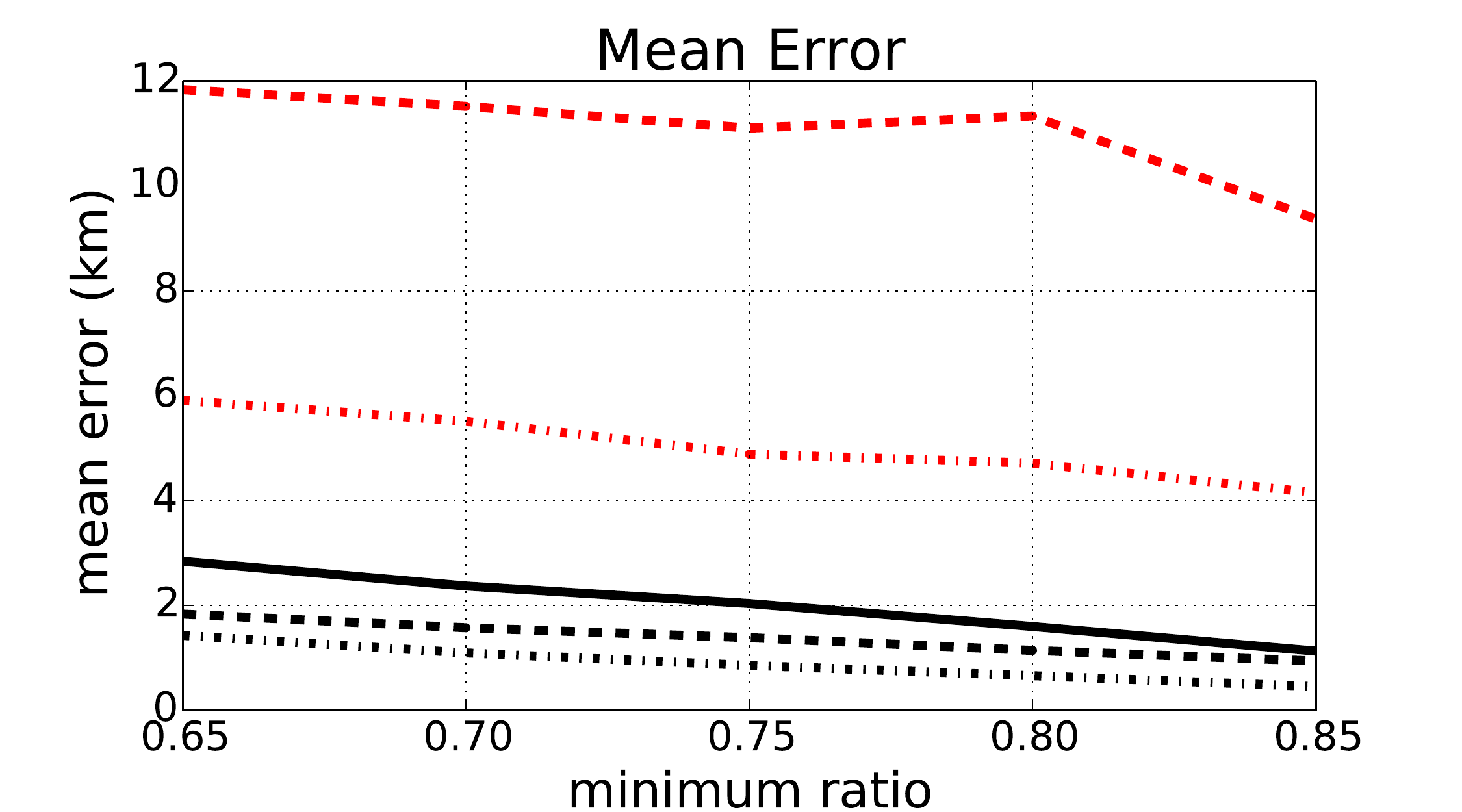}}
\subfigure{\includegraphics[width=0.8\columnwidth]{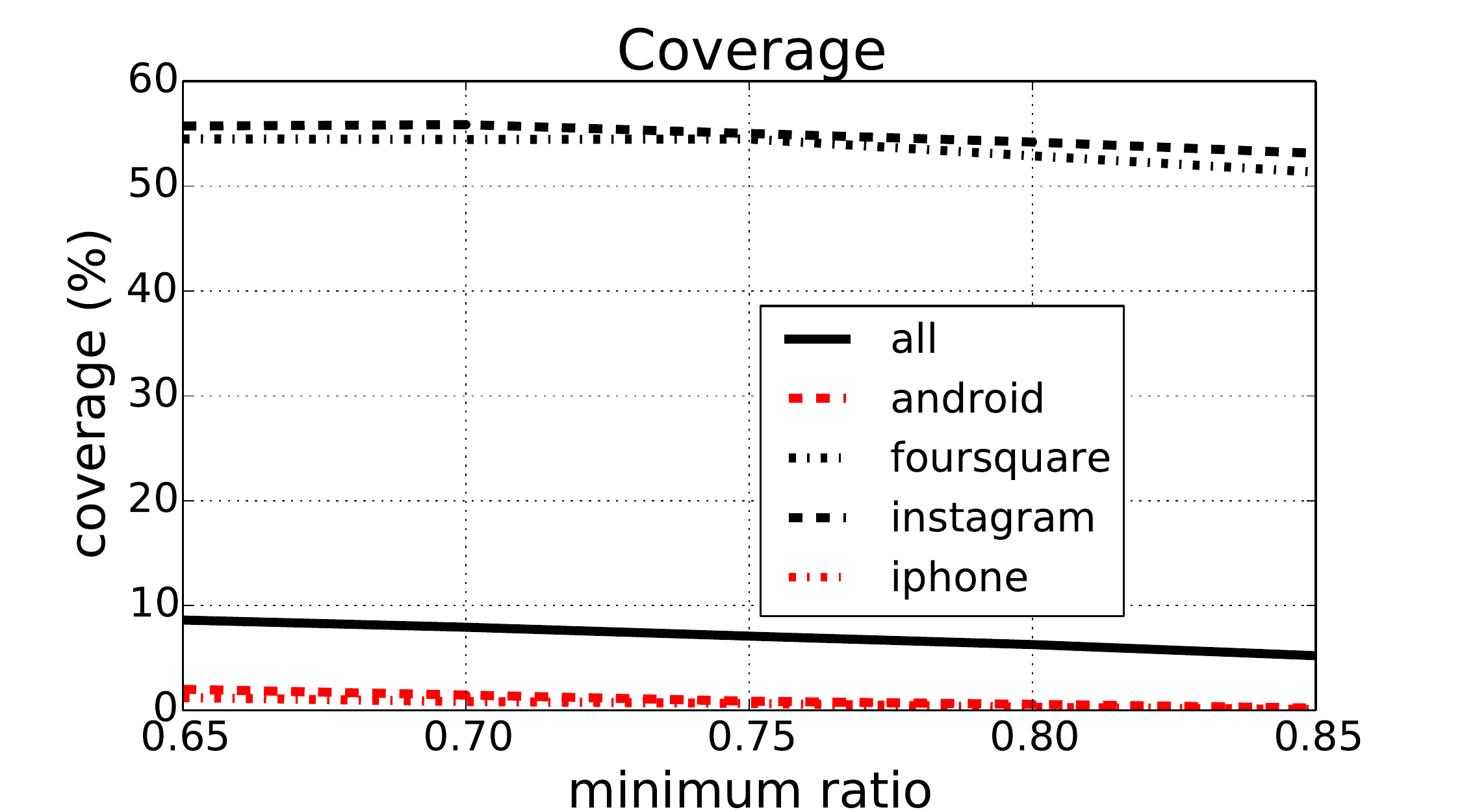}}
\subfigure{\includegraphics[width=0.8\columnwidth]{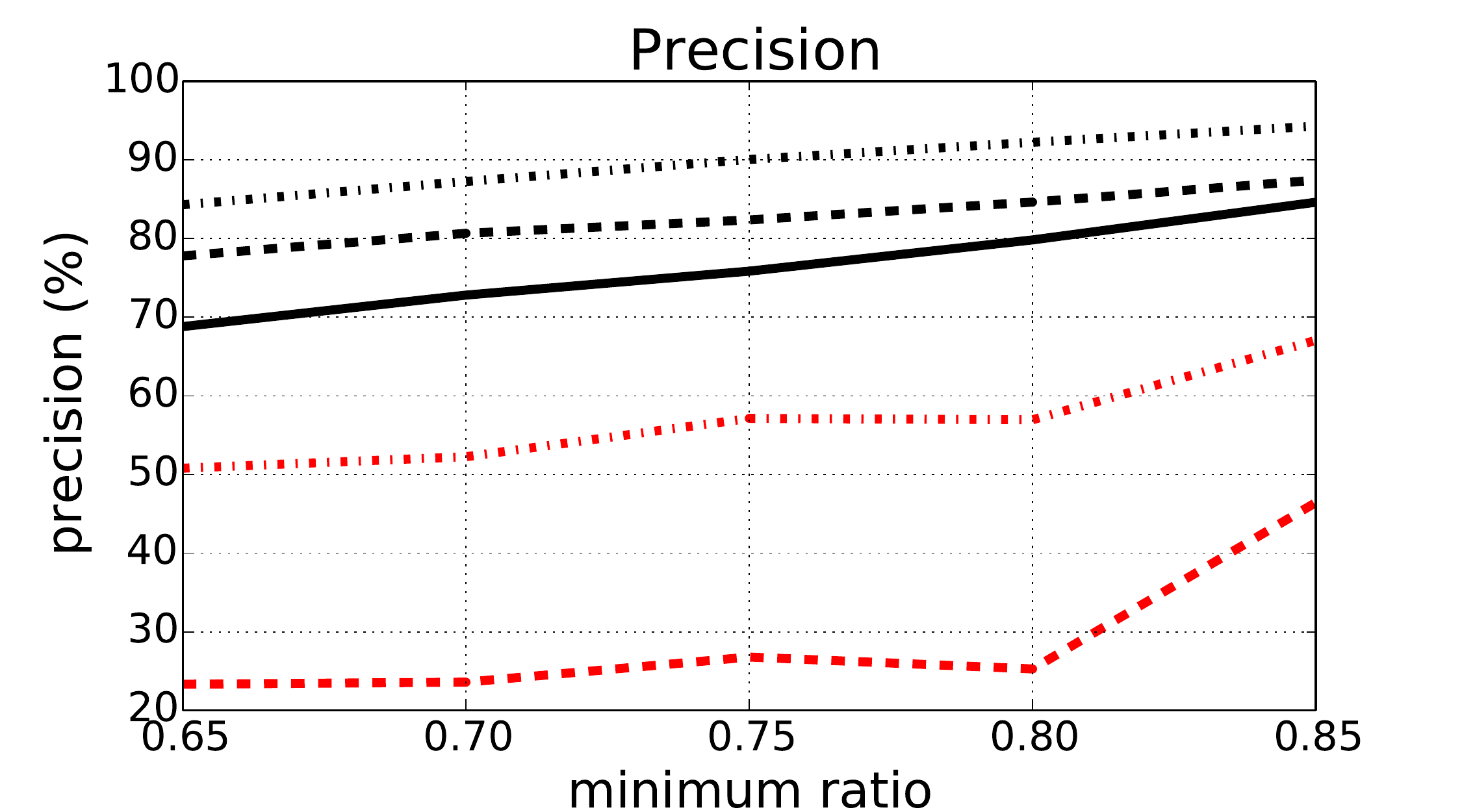}}
\subfigure{\includegraphics[width=0.8\columnwidth]{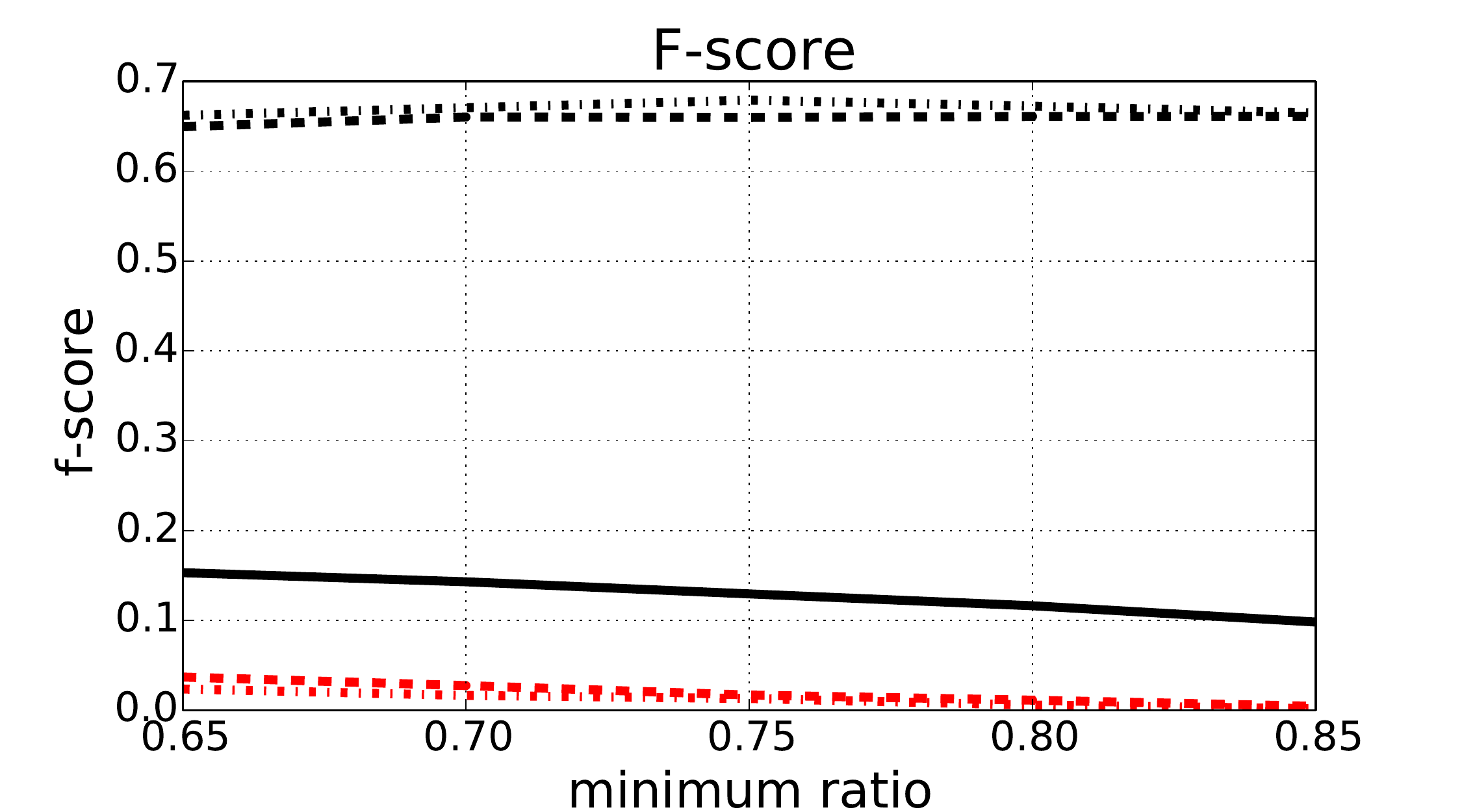}}
\caption{Effects of varying $\tau$ (minimum ratio) on performance for different datasets fixing $s = 4km^2$ }\label{figure_params_ratio}
\end{center}
\end{figure}
 
\begin{figure}[h!]
\begin{center}
\subfigure{\includegraphics[width=0.8\columnwidth]{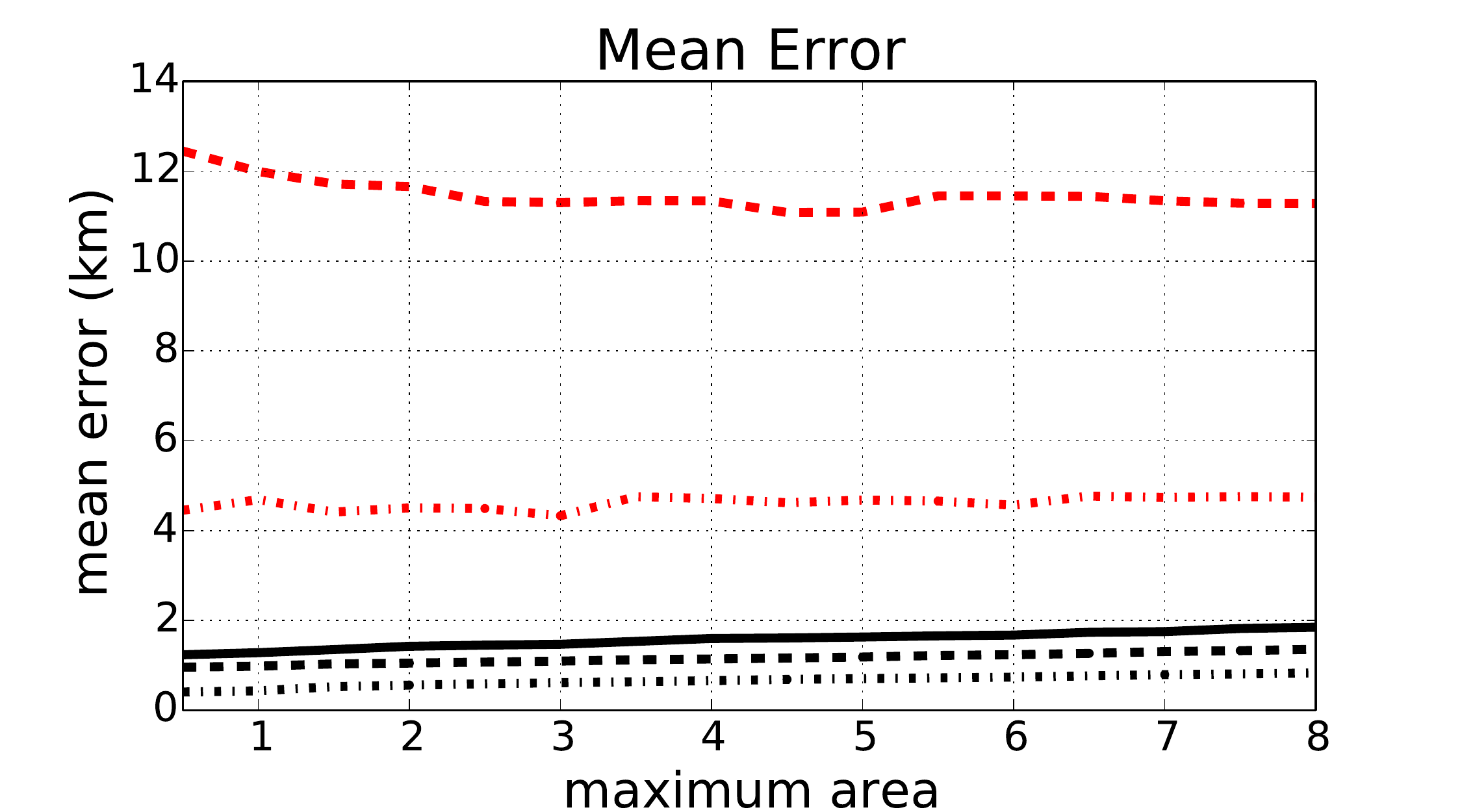}}
\subfigure{\includegraphics[width=0.8\columnwidth]{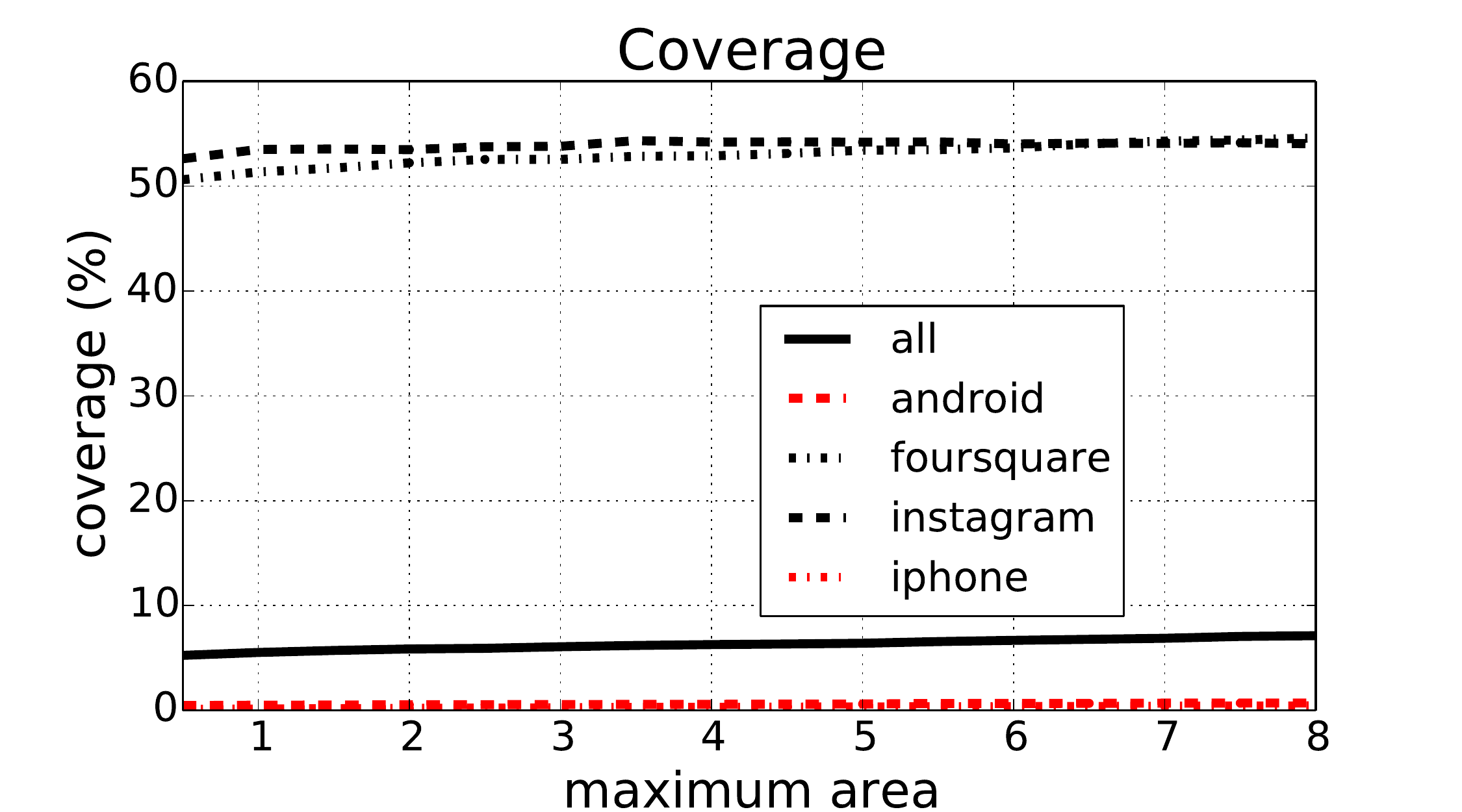}}
\subfigure{\includegraphics[width=0.8\columnwidth]{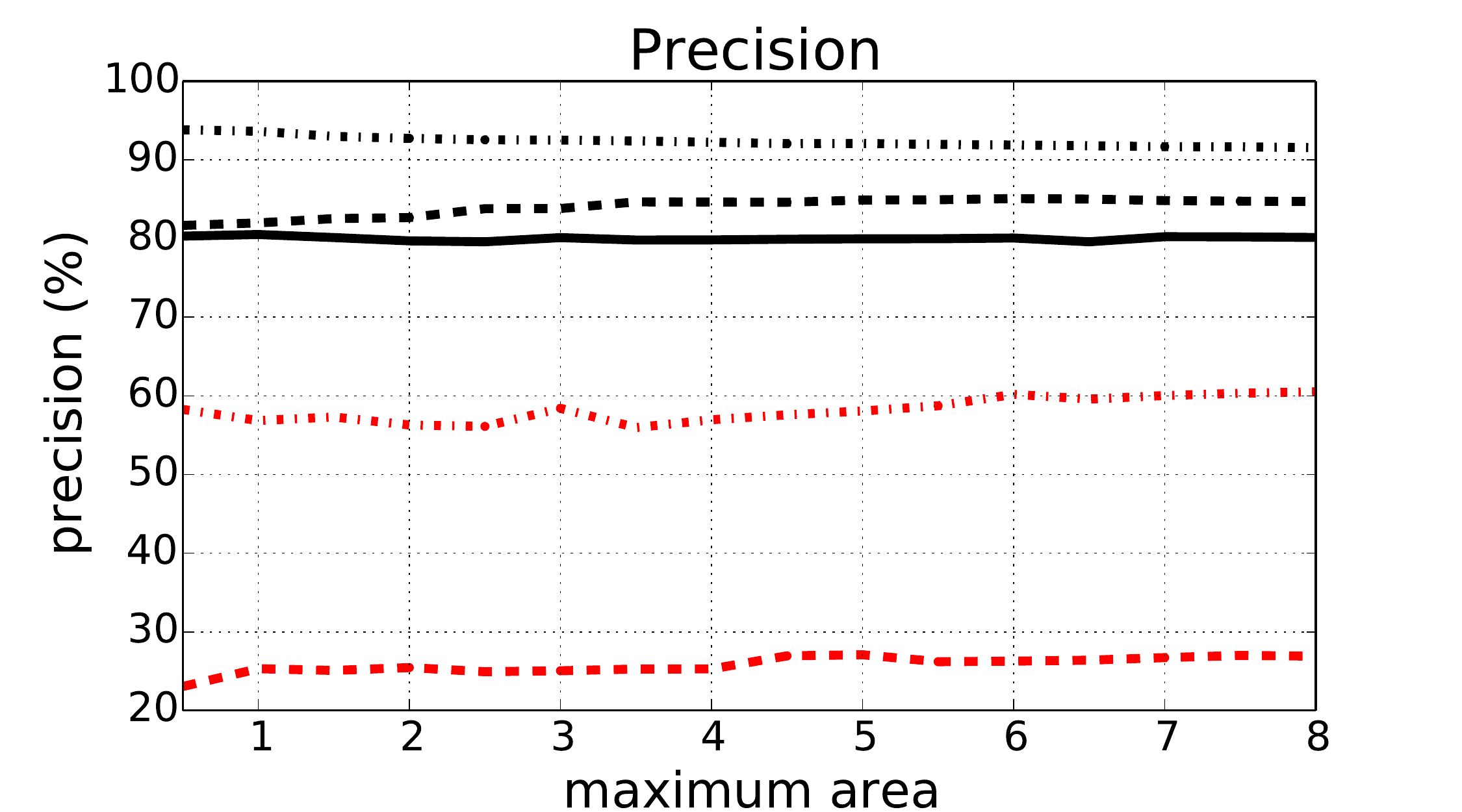}}
\subfigure{\includegraphics[width=0.8\columnwidth]{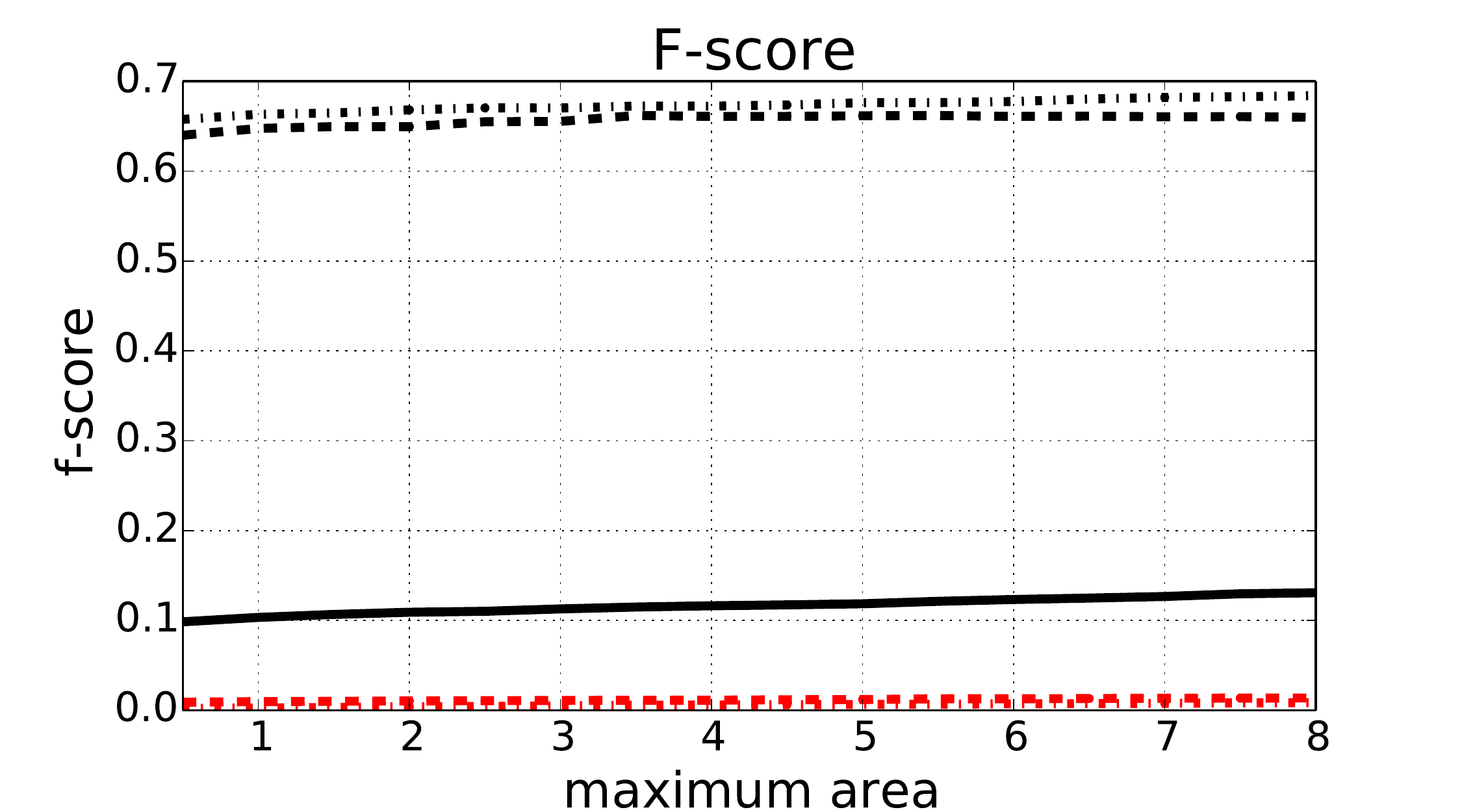}}
\caption{Effects of varying $s$ (maximum area) on performance for different datasets fixing $\tau = 0.8$}\label{figure_params_area} 
\end{center}
\end{figure}

The results are shown in Figures~\ref{figure_params_ratio} and~\ref{figure_params_area}. Figure~\ref{figure_params_ratio} shows results for varying the minimum ratio $\tau$ with maximum allowed area $s$ fixed at $4km^2$. Figure~\ref{figure_params_area} varies $s$ with the minimum ratio $\tau$ fixed at $0.8$. The curves shown are for all the different datasets, where each curve represents the results for training and testing on the same source. For example, the  solid black lines in Figure~\ref{figure_params_ratio} show that when using TW-All for training (using the training sample) and testing (on the test data), when the minimum ratio is set to $0.8$, the mean error is $1.6$, the coverage is 6.3\% (6.3\% of tweets in the test set can be assigned a location), the precision is 79.8\%, and the F-score is $0.12$ (dragged down by the low coverage). 

While the results seem promising, there is significant variation in performance between the datasets. In particular, the performance for geotagging content from Instagram and Foursquare demonstrates high accuracy, precision and coverage. At the same time, results for predicting locations for the iPhone or Android datasets are low in all these metrics, and especially in terms of coverage. For the TW-All dataset, as a random sample from the content (and with a heavy representation of iPhone data), results are in between. 

In reality, we can have the most impact for real-world applications by geotagging Android/iPhone and Instagram content. Content originating from Foursquare is already geotagged: if not associated with location on the Twitter platform, it includes a link to a Foursquare venue that pinpoints the content by identifying the venue. Instagram content, however, may be more interesting to fill locations for --- we provide some preliminary numbers below in Section~\ref{sec_heatmap}. Most important, though, are tweets from Android and iPhone, as (1)~they represent a much larger portion of Twitter data, (2)~they are less likely to be geotagged (see Section~\ref{sec_heatmap}), and (3)~they are most similar in characteristics to other sources of Twitter items (e.g. the Twitter web site  \cite{Priedhorsky_2014}).


Figure~\ref{figure_params_area} also demonstrates that the results are not very sensitive to  $s$, the maximum area size parameter. In other words, when an $n$-gram demonstrates strong geo-specific tendencies, the model correctly captures and represents that information regardless of area size.

Next, we explore the performance based on specific test sets, focusing on TW-iPhone and TW-Instagram, while experimenting with different \emph{training} sets. 

\subsection{Cross-Model Performance}\label{sec_cross_model}

\begin{figure}[t]
\begin{center}
\subfigure[test set = TW-iPhone \label{figure_cross_model_iphone}]{\includegraphics[width=0.9\columnwidth]{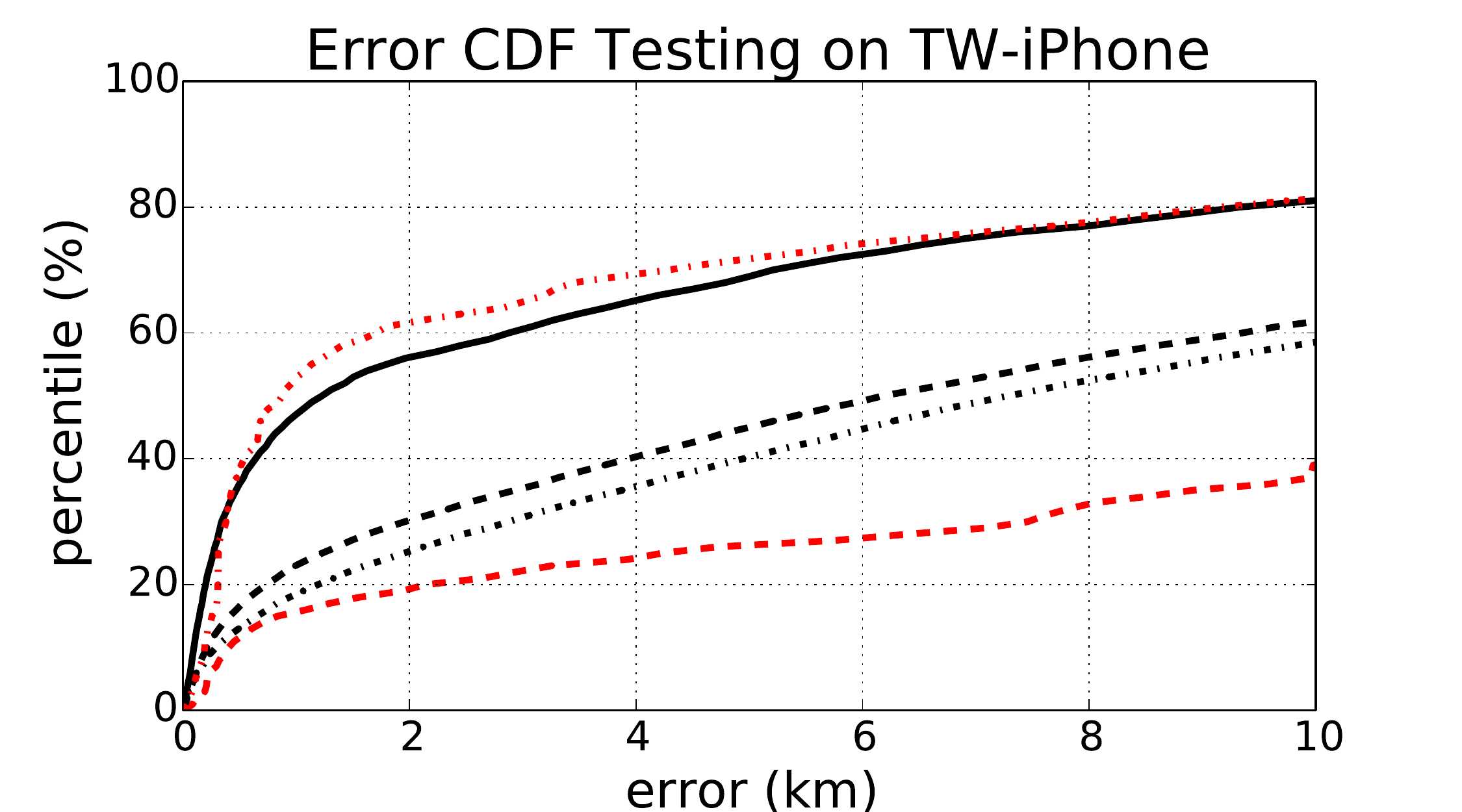}}
\subfigure[test set = TW-Instagram\label{figure_cross_model_instagram}]{\includegraphics[width=0.9\columnwidth]{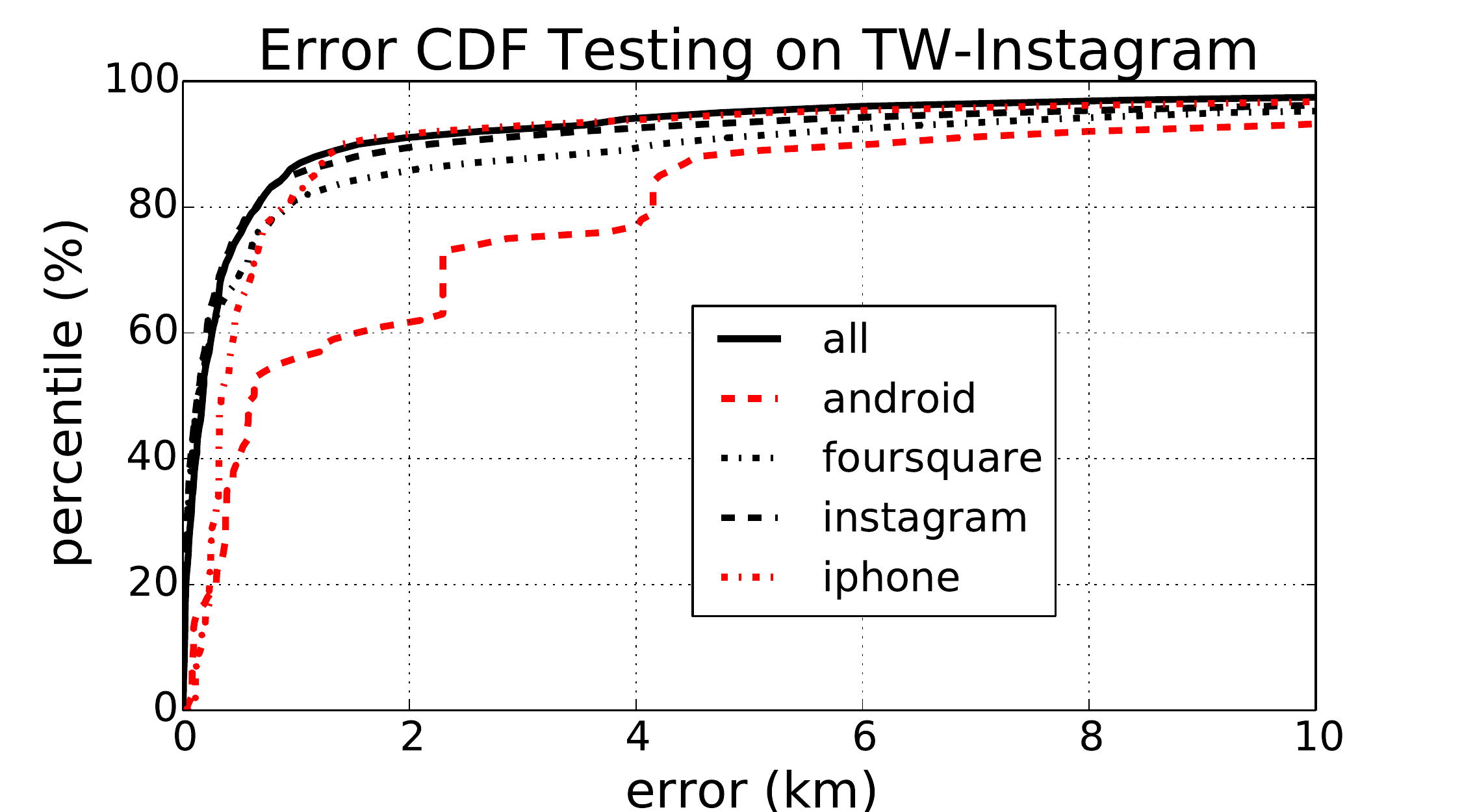}}
\caption{Performance when training with different datasets and testing on TW-iPhone and TW-Instagram}\label{figure_cross_model}
\end{center}
\end{figure}

Figure~\ref{figure_cross_model_iphone} shows the performance when training on different datasets, while testing on the key TW-iPhone data set (i.e., on tweets posted from the Twitter for iPhone application). The $n$-gram extraction and modeling was performed on multiple training sets with parameters $s=4km, \tau=0.8$. One can see in Figure~\ref{figure_cross_model_iphone} that when using the TW-All dataset for training, roughly half of the geotagged tweets in the TW-iPhone test data set are geotagged within a $1.0km$ radius of their true location. Somewhat surprisingly, the accuracy of the results increases when using TW-All compared to using TW-iPhone for training (recall that TW-All is a sample of all tweets, with roughly 60\% of tweets emerging from iPhone). We believe that the reason for the improved accuracy is the additional coverage provided by $n$-grams that are detected as location-specific with support from Foursquare and Instagram. 

There are significant differences in coverage between the training sets. Using the TW-iPhone training set results in much lower coverage (0.3\%) than TW-All (1.3\%), TW-Foursquare (6.4\%) and TW-Instagram (5.1\%), when testing on TW-iPhone. 
One possible explanation for this low coverage, which we discuss more in Section \ref{sec_concl}, is the weak \textit{aboutness} of TW-iPhone content and  lower rate of location references, relative to other datasets such as TW-Foursquare or TW-Instagram. Our experiments support this hypothesis due to the fact that the low coverage is driven not by the failure to locate geo-specific $n$-grams but rather by the failure to identify a large number of geo-specific $n$-grams.

Figure~\ref{figure_cross_model_instagram} shows the same analysis when testing on the TW-Instagram dataset. Clearly, the results are significantly better than testing on TW-iPhone with more than 80\% of the items geotagged within $1.0km$ of their true location when using datasets other than TW-Android for training. The coverage is also much better across the board: 38.1\% for training on TW-All, 47.2\% for TW-Foursquare, 54.2\% for TW-Instagram, 3.8\% for TW-Android, and 4.6\% for TW-iPhone.

In summary, the results of our cross-model experimentation show that when geotagging hyper-local content for search and data mining apps, not all data is created equal. Considering the source of the training data, as well as the source of the test data could prove critical to the performance and accuracy of the solution. Next, we explore the differences in location models that are created based on different sources of data.  





\section{n-gram "Gravity"}\label{sec_gravity}
\begin{figure}[h!]
\begin{center}
\includegraphics[width=0.9\columnwidth]{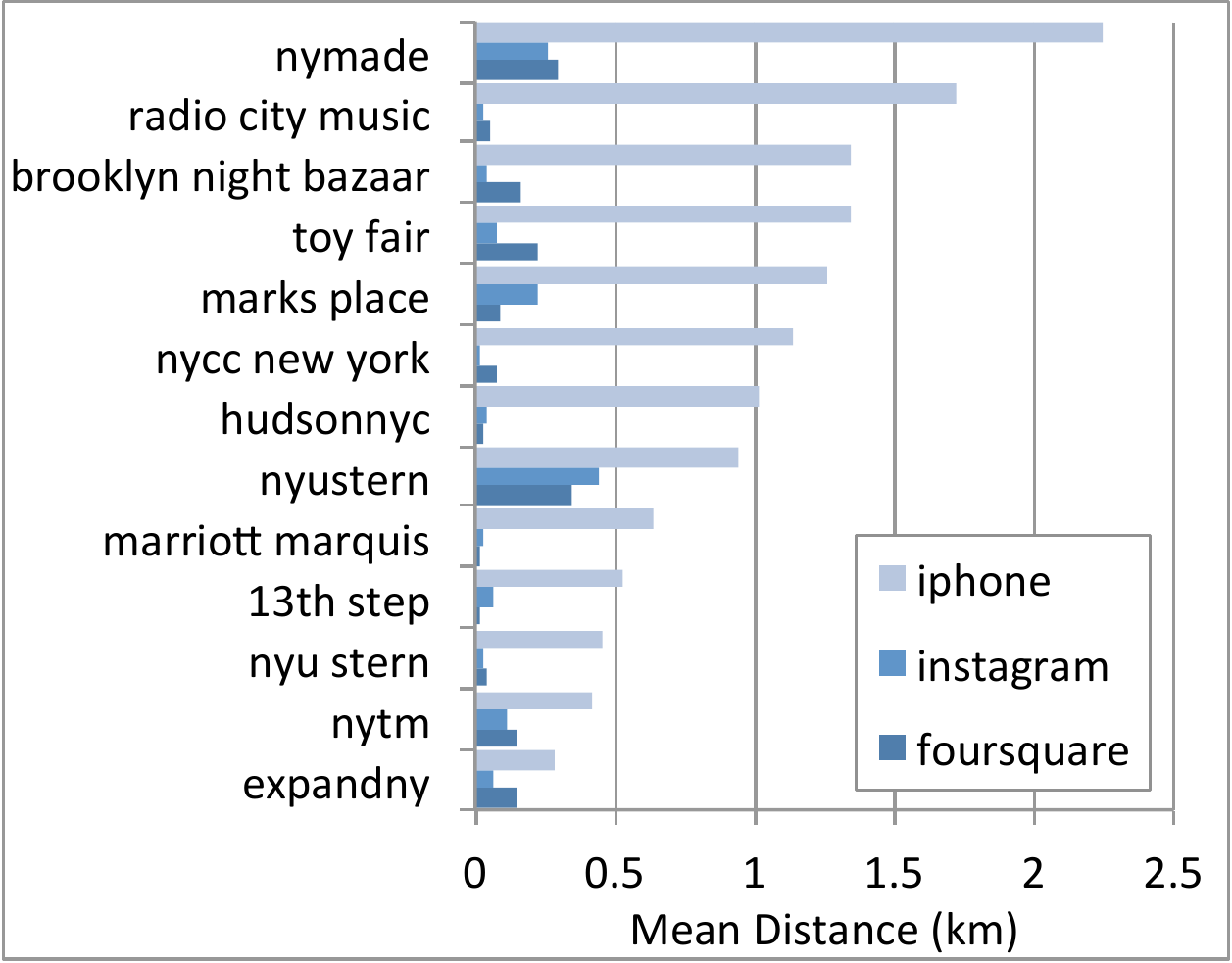}
\caption{N-Gram gravity for multiple sources}\label{figure_gravity}
\end{center}
\end{figure}

As we have seen in previous sections, the performance in terms of accuracy and precision varies greatly between the different sources. Are there significant differences between location models for $n$-grams that are geo-specific across different sources, or are the differences only due to different $n$-grams are extracted from each training dataset? In this section we touch on the differences in ``gravity'', our colloquial term referencing the dispersion of tweet locations, for $n$-grams that are detected as geo-specific across three different training datasets: TW-iPhone, TW-Instagram and TW-Foursquare. Figure~\ref{figure_gravity} shows that there are significant differences in the dispersion of locations for the three datasets, even for the same $n$-grams. The figure shows, for the top thirteen $n$-grams that were identified as geo-specific in each of the three sources, the average distance between the location $l_i$ of tweet in the \textit{training} set, to the center of the final Gaussian model $\mathcal{N}_j$. For example, for the $n$-gram ``NYCC New York'', in the TW-iPhone dataset, the average distance between the $n$-gram tweets in the training set to the center of the model was $1.14km$. The same distance computed on the TW-Instagram training set was $0.015km$.

There are multiple potential contributors to the fact that the iPhone data is more dispersed, including the ``aboutnesss'' of content, the accuracy of location provided by the application, and the semantics of posts from the different sources. We discuss these in more length in Section~\ref{sec_concl}. Regardless of the reason for the differences in dispersion, it is clear that the phenomena has a direct effect on the results for the different datasets.

\section{Potential Gains}\label{sec_heatmap}
We have seen in Section ~\ref{sec_gravity} that we can achieve a significant gain in items that are associated with a precise location, at least for certain types of data and items. Under these assumptions, what are the patterns and scale of these gains?

We measure the relative frequency of posts geotagged in the New York area coming from different application sources in Table~\ref{table_datasets}. Ideally we would like to geotag every post that has hyper-local geographic relevance. But what fraction of posts, in the best case scenario, can be geotagged? Twitter is reported to have about 2\% geotagged content, but there are no reported numbers of portions of geotagged for individual sources. We performed a preliminary study where we track a sample set of keywords on the Twitter stream to estimate the portion of geotagged items posted to Twitter by the different applications. For example, between 1\% and 10\% of the tweets posted from Twitter for iPhone for the keywords we tracked where geotagged (3\% for the keyword ``New York'', 4\% for the common word ``at''). Foursquare showed a much higher ratio, anywhere from 22\% to 72\% (26\% for ``New York'', 71\% for ``at''). Posts made from Twitter for Android ranged between 2\% and 17\% for the different terms (2\% for ``New York'', 4\% for ``at'') and Instagram posts to Twitter were, for the phrases we tracked, between 10\% and 42\% geotagged (``New York'': 42\%, ``at'': 14\%).

\begin{figure}[h!]
\begin{center}
\subfigure[training data locations\label{figure_heatmap_train}]{\includegraphics[width=0.9\columnwidth]{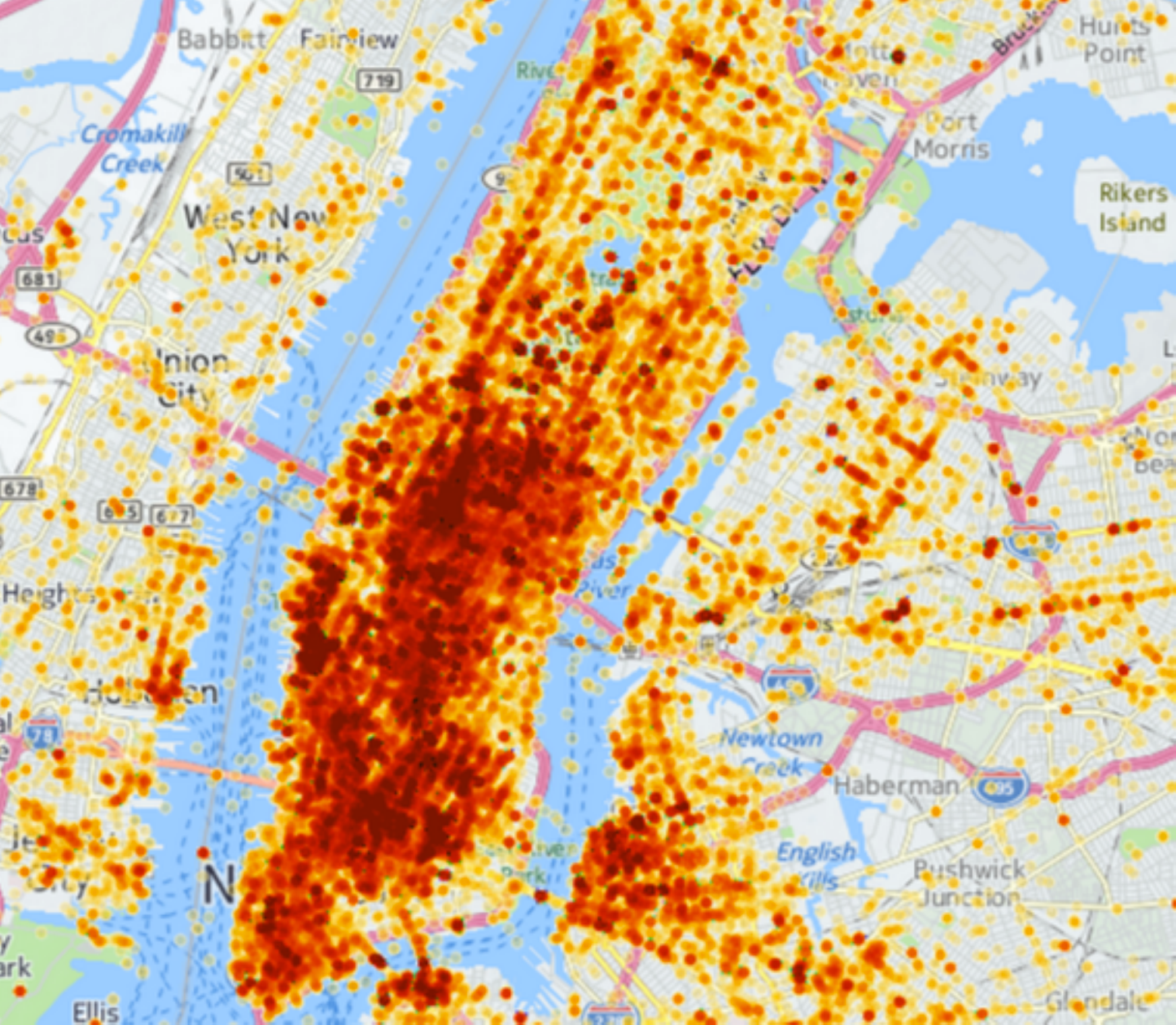}}
\subfigure[predicted locations (gain)\label{figure_heatmap_pred}]{\includegraphics[width=0.9\columnwidth]{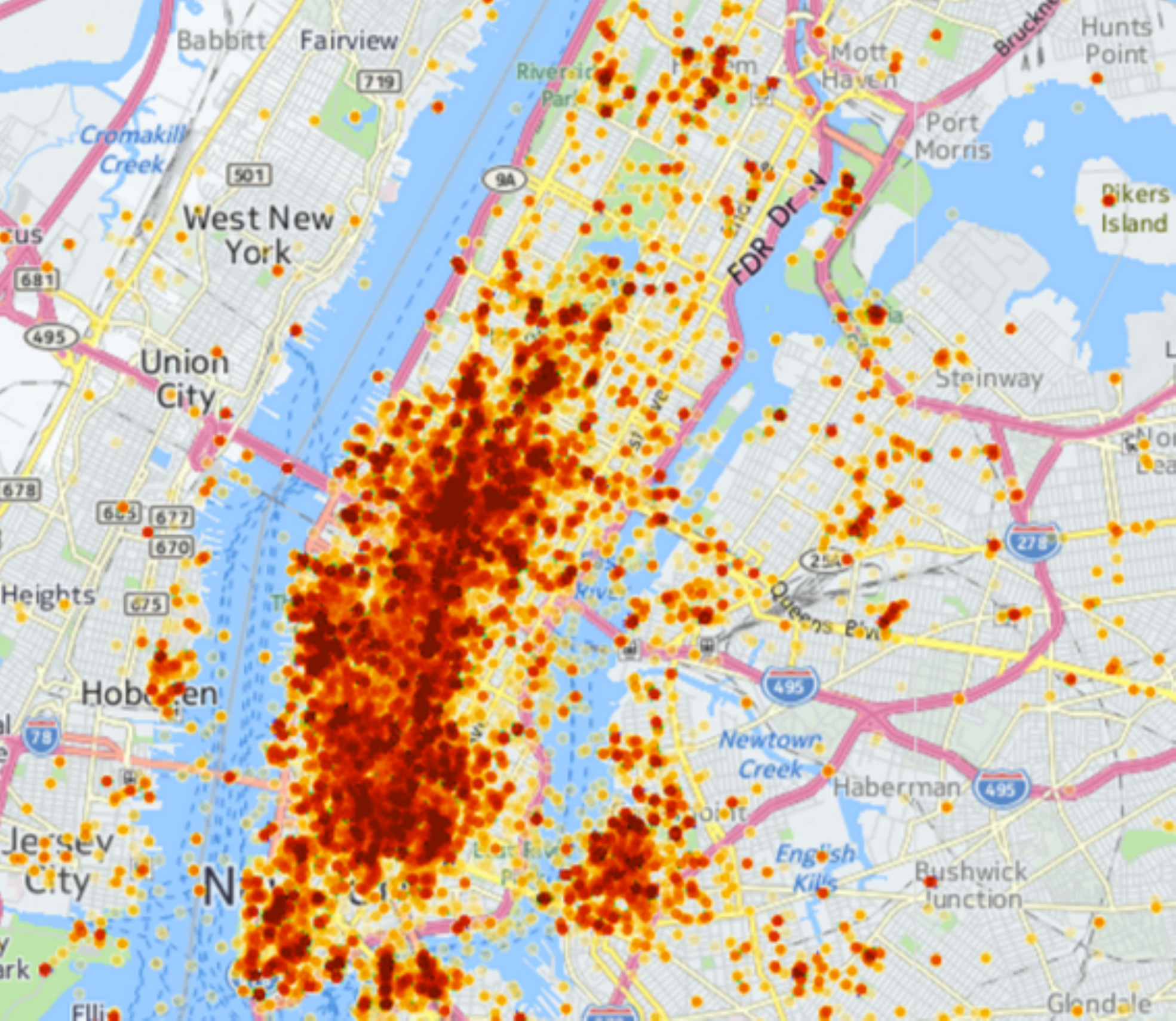}}
\caption{Heatmaps of training and predicted tweets from TW-Instagram dataset, sampled to use same number of observations}\label{figure_heatmap}
\end{center}
\end{figure}

These numbers, with our results above, indicate an opportunity for significant gains in the amount of content associated with hyper-local data. For example, for content from Instagram, we report 60\% coverage for our default parameter settings (see Figure~\ref{figure_params_ratio}). In other words, if just 20\% of the Instagram content posted to Twitter is geotagged, we can get a further lift of $0.6 \times 0.8 = 0.48$ of the items posted from Instagram, more than tripling the amount of available data. This potential boost is performed with precision close to 90\% (Figure~\ref{figure_params_ratio}).

How is this gain distributed? Figure~\ref{figure_heatmap} shows the geographic distribution of content items in the TW-Instagram training set (Figure~\ref{figure_heatmap_train}) and the locations of items associated with predicted location from the test set (Figure~\ref{figure_heatmap_pred}) (we randomly sub-sampled from the training set so both figures show the same number of items). We used the default settings for the analysis in this figure ($s=4.0km, \tau=0.8$), and show only a portion of the geographic area we covered in our analysis. Points that are colored red indicate a higher concentration for that area. As is evident in the figure, most of the gain occurs in the already-popular areas where most content is posted (i.e., midtown and downtown Manhattan). Indeed, the methods we propose above bring a ``rich get richer'' phenomena, where the models are most robust, and allow for associating content with additional social media items, in the areas that are already covered.

\section{Discussion and Conclusions}\label{sec_concl}
A data-driven approach for geocoding individual social media items at a hyper-local scale has the potential to extend the geographic coverage of social media data, but its performance may depend on the distribution of the data according to its source. We used Twitter data to identify and model geo-specific $n$-grams, based on the location distribution of tweets associated with them. We then predicted locations for individual tweets based on a given tweet's geo-specific $n$-gram's locations (if any). The performance of this method was highly contingent on the source of the data. Data from the check-in application Foursquare and the photo sharing application Instagram were highly location-specific in general, and as a result, the method produced location estimates with high accuracy. Conversely, tweets from ``regular'' Twitter clients, like Twitter for iPhone, demonstrated low accuracy, and even when geo-specific $n$-grams were detected from these sources, the breadth of the location model for these $n$-grams was much larger than the same model in data posted to Twitter through Instagram or Foursquare.

There are a number of possible reasons for the differences between sources, chief among them the \textit{aboutness} of the content and the density of location-based references. We regard the aboutness of posts in social media as the likelihood of the posts to be about a geo-specific feature. Foursquare has clear and strong aboutness as posts are often of the format ``I am at\ldots''. Instagram, as a photo sharing service, also has strong aboutness: when a photo is taken, the text is very likely to reference to object that is in it, and that object is likely to be nearby (although it may not be, e.g. in the case of the Empire State Building). For a regular Twitter client, aboutness could be more dispersed, as people are more likely to make comments like ``I am headed to Central Park'' or ``I hear the Empire State Building was lit in blue and white today'' --- both comments that are quite unlikely on Foursquare and Instagram. Related, the density of location references is lower in non-Foursquare and non-Instagram content, as people may converse on any topic.


Nevertheless, we have shown that significant reach can be achieved for certain applications under certain conditions with a potential to more than triple the amount of available, geotagged content for Instagram, for example. However, the lift that is gained demonstrates the ``rich get richer'' phenomena, where places that were already significantly covered are more likely to gain new content than other areas. This phenomena is likely to bias  the distribution of social media geotagged content even more significantly.  Hecht and Stephens describe the implications of such bias in respect to urban and rural communities \cite{hecht_2014}.

It is important to note that the results described here are somewhat over-optimistic because they are based on data from a single locale (i.e. New York City). For instance, an $n$-gram like ``city hall'' may be geo-specific when examining a New York-only dataset, but in a world-wide dataset there would be many different city halls. However, the concern can be somewhat mitigated by additional information a system may have, e.g. the user profile information or the IP address.


Another key limitation in our approach is the bias introduced by building models based on geotagged data. In fact, Twitter had recently reported (see footnote~1) that 1\% of the users produce 66\% of the geotagged items. It is entirely possible that the data we trained and tested on is significantly biased, e.g., the type of content that is posted may be different between users that geotag their content versus others. Of course, bias can also result from the fact that most geotagged content is posted from specific Twitter applications. These concerns are also significant, but can be mitigated by a couple of factors. First, as we show above, we can achieve significant lift even when training and testing with the data from the same source, somewhat limiting the bias. Second, as reported by Priedhorsky et al.~\cite{Priedhorsky_2014}, the difference between geotagged and non-geotagged tweets is limited; the authors report a correlation of 0.85 between the unigram frequency vectors for each set. On the other hand, there are a number of avenues for future work that could further improve the results reported here. First, using language analysis and, in particular, extracting tense and time references \cite{Li_2014} could help improve the models for TW-iPhone data where people are likely to post plans and reviews with their location references. A more refined approach to combining multiple $n$-grams that appear in tweets could take into account their likelihood and co-variation models. Additionally, using user level information, e.g. profile or historic data, has the potential to greatly improve performance. Further, our model could be refined to not only rely on geographic information, but include temporal information as well \cite{Rattenbury_2007}. Note, however, that a photo's geographic and temporal metadata may not always be accurate \cite{Moreno_2014}. Nonetheless, modeling by space, time and source may allow more gains in geolocating content in hyper-local areas.

Ultimately, the gain in content associated with hyper-local geographic areas can help create better models, and understand better the activities in different geographic areas. For example, we can use these data to provide more robust information to people interested in monitoring the activities, say, in Central Park. Such gains will provide for new ways to reflect the activities in the park, detect outliers and unusual activities, and help users with specific search and data mining tasks.

 \section{Acknowledgements}

This work is supported in part by the National Science Foundation grant numbers 1446374 and 1444493. Any opinions, findings, and conclusions or recommendations expressed in this material are those of the author(s) and do not necessarily reflect the views of the National Science Foundation. The work of Yaron Kanza was supported in part by ISF grant 1467/13. Yana Volkovich was supported by the People Programme (Marie Curie Actions, from the FP7/2007-2013) under grant agreement no.600388 managed by REA and ACCI\'O.







\begin{spacing}{0.92}
\bibliography{bibliography/converted_to_latex}
\end{spacing}

\end{document}